\documentclass[preprint,12pt,amsmath,amssymb]{revtex4-1}
\usepackage{ifthen}
\usepackage{graphicx}
\usepackage{bm}
\newboolean{FIG}
\setboolean{FIG}{true}
\usepackage[usenames]{color}

\usepackage{hyperref}
\begin{document}
\def \tr{{\mbox{tr~}}}
\def \ra{{\rightarrow}}
\def \ua{{\uparrow}}
\def \da{{\downarrow}}
\def \be{\begin{equation}}
\def \ee{\end{equation}}
\def \ba{\begin{array}}
\def \ea{\end{array}}
\def \bea{\begin{eqnarray}}
\def \eea{\end{eqnarray}}
\def \nn{\nonumber}
\def \l{\left}
\def \r{\right}
\def \half{{1\over 2}}
\def \etal{{\it {et al}}}
\def \cH{{\cal{H}}}
\def \cM{{\cal{M}}}
\def \cN{{\cal{N}}}
\def \cQ{{\cal Q}}
\def \cI{{\cal I}}
\def \cV{{\cal V}}
\def \cG{{\cal G}}
\def \cF{{\cal F}}
\def \cZ{{\cal Z}}
\def \bS{{\bf S}}
\def \bI{{\bf I}}
\def \bL{{\bf L}}
\def \bG{{\bf G}}
\def \bQ{{\bf Q}}
\def \bR{{\bf R}}
\def \br{{\bf r}}
\def \bu{{\bf u}}
\def \bq{{\bf q}}
\def \bk{{\bf k}}
\def \bz{{\bf z}}
\def \bx{{\bf x}}
\def \ttau{\tilde{ \tau}}
\def \tS{\tilde{ \Sigma}}
\def \hK{\hat{ K}}
\def \bpsi{{\bar{\psi}}}
\def \tJ{{\tilde{J}}}
\def \W{{\Omega}}
\def \e{{\epsilon}}
\def \lam{{\lambda}}
\def \L{{\Lambda}}
\def \S{{\Sigma}}
\def \a{{\alpha}}
\def \t{{\theta}}
\def \b{{\beta}}
\def \g{{\gamma}}
\def \D{{\Delta}}
\def \d{{\delta}}
\def \w{{\omega}}
\def \s{{\sigma}}
\def \f{{\varphi}}
\def \x{{\chi}}
\def \e{{\epsilon}}
\def \h{{\eta}}
\def \G{{\Gamma}}
\def \z{{\zeta}}
\def \hatt{{\hat{\t}}}
\def \hn{{\bar{n}}}
\def \ts{{\tilde{\sigma}}}
\def \vk{{\bf{k}}}
\def \vq{{\bf{q}}}
\def \gk{{\g_{\vk}}}
\def \nd{{^{\vphantom{\dagger}}}}
\def \yd{^\dagger}
\def \av#1{{\langle#1\rangle}}
\def \ket#1{{\,|\,#1\,\rangle\,}}
\def \bra#1{{\,\langle\,#1\,|\,}}
\def \braket#1#2{{\,\langle\,#1\,|\,#2\,\rangle\,}}
\newcommand{\kg}{{kagome }}

\setcounter{secnumdepth}{3}

\title{Universal dynamics and renormalization in many body localized systems}
\author{Ehud Altman and Ronen Vosk}
\affiliation{Department of Condensed Matter Physics, Weizmann Institute of Science, Rehovot 76100, Israel}
\begin{abstract}
We survey the recent progress made in understanding non equilibrium dynamics in closed random systems. The emphasis is on the important role played by concepts from quantum information theory and on the application of systematic renormalization group methods to capture universal aspects of the dynamics. Finally we outline some outstanding open questions, which includee the description of the many-body localization phase transition and identifying physical systems that will allow systematic experimental study of these phenomena.
\end{abstract}
\maketitle

\tableofcontents

\section{Introduction}

The rapid development of experiments and increasing complexity of the physics observed in systems of ultra-cold atomic gases\cite{Bloch-review}, which are essentially decoupled from an external heat bath, has lead to a surge of interest in the dynamics of closed quantum systems\cite{Polkovnikov-review,Lamacraft-Moore-Review}.
There is a common expectation that complex many-body systems with sufficiently generic interactions will ultimately attain thermal equilibrium when left to evolve on their own. In the process, any information that had once existed in the initial state is lost, except information on global conserved quantities, such as the energy or particle number. In particular, quantum information that might have been stored in local degrees of freedom rapidly becomes irretrievable as these degrees of freedom become increasingly entangled with a macroscopic number of other particles. The last structures of information to survive in the system before it sinks to the boring homogenous reality of thermal equilibrium are the long wavelength modes, which describe the transport of global conserved quantities across the system and the dynamics of coarse grained order parameter fields.  This leads to a classical description of the dynamics in thermalizing systems that has been immensely successful for a broad range of applications\cite{Hohenberg1977}.

However, not all physical systems must thermalize.
The premise underlying the work we discuss in this review is that quantum systems of interacting particles subject to sufficiently strong disorder
will fail to come to thermal equilibrium by themselves when they are not coupled to an external bath. This phenomenon is commonly referred to as many-body localization.
In absence of  thermalization
quantum information stored in local degrees of freedom can persist and influence the dynamics at long times. Therefore a description of the dynamics in such systems, even in the long time limit, requires a quantum theory. An  effective classical theory formulated in terms of stochastic equations of motion as in Ref. [\onlinecite{Hohenberg1977}] will not suffice to describe the dynamics.
The goal of this review is to survey and explain the recent progress made toward understanding the quantum
dynamics in the non-thermalizing phase. Notions of quantum information theory, such as entanglement and its propagation in the system, play a central role in the emerging theoretical descriptions as do renormalization group ideas which allow to capture the universal features of the dynamics.

First, let us briefly recall the pre-history of the many-body localization problem and recount how it evolved into the questions that are of current interest in the field.
The idea that certain closed random systems may fail to thermalize was first raised by Anderson in his  famous paper on localization\cite{Anderson1958}. There, Anderson actually solved for the propagation of a single particle in a random potential, but he was motivated by a question on many-body dynamics: under what conditions spin or energy diffusion will take place in a closed system of interacting localized spins? In this context the single particle problem Anderson actually solved can be viewed as the propagation of a single up spin in the background of all down spins. By contrast the many-body problem concerns the time evolution starting from an initial state with a more generic configuration of interacting spins. For example, suppose we prepare the system in the following spin configuration $\ket{\psi_i}=\ket{\ua\ua\da\ua\da\da\da\dots}$.
Then we let it evolve under the influence of a Hamiltonian with interactions and random local magnetic fields and ask if and how the system reaches equilibrium at long times. Specifically we could ask if the expectation value of a local spin operator $\av{S^z_i}$ relaxes to its equilibrium value. Note that the time evolution can in principle involve (when symmetry is minimal) up to $2^N$ other basis states, making it immensely more complex than the propagation of a single flipped spin. Localization is therefore by no means obvious a priori even if all single particle states are localized.

An alternative perspective on the same problem is gained from considering eigenstates. Eigenstates of thermalizing systems are expected to obey the eigenstate thermalization hypothesis (ETH)\cite{Deutsch1991,Srednicki1994,Rigol2008}. This hypothesis can be viewed as an application of the equivalence of ensembles to the extreme limit in which the micro canonical energy window includes only one eigenstate. The ETH then asserts that any reasonably local correlation function measured in an energy eigenstate of a closed system should appear as a thermal correlation function with the appropriate temperature. As a diagnostic of localization we may ask if the eigenstates of the interacting spin problem violate ETH.
Specifically, they may be smoothly connected, in a sense that will be made more precise later, to localized (product) states of spins. We shall later see, however, that there are non-thermal localized states that nonetheless cannot be smoothly connected to a product state. Such states will have a more general matrix-product or tensor network representation. In any case, the Von-Neuman entropy associated with a subsystem, when the full system is in a localized state, scales like the area of the sub-system. {\em Such area-law scaling of the entanglement in energy eigenstates of finite energy density can in fact be taken as a defining property of the localized phase} \cite{Bauer2013}.
This should be contrasted with thermal eigenstates which would show volume law scaling of the entanglement entropy.
From the above discussion it should be clear that the generic and also more interesting setting for many-body localization involves eigenstates with an extensive energy above the ground state.
We shall see that understanding the structure of high energy eigenstates can be a powerful theoretical tool in analyzing the dynamical properties of the system.

Of course similar questions, regarding time evolution or the structure of eigenstates,  concern not only spin systems but also systems of fermionic or bosonic particles. Localized eigenstates of noninteracting particles can be constructed as  appropriately symmetrized (or anti-symmetrized) wave functions of localized single particle filled-states. We may then ask what is the fate of these states when interactions are turned on. Stability of a localized state is not guaranteed a priori as it can potentially delocalize by coupling resonantly to exponentially many other localized states through high order interaction processes.

Convincing perturbative arguments that the localization, even in high energy states, is indeed stable to interactions were given a few years ago in Refs. \cite{Gornyi2005,Basko2006}. These arguments are nicely summarized by another review paper in the same volume\cite{Nandkishore2014}. Basko et. al \cite{Basko2006} have emphasized that the many-body localized state can be viewed as a dynamical phase and that a phase transition between a localized and delocalized phase can be tuned by varying the energy density of the system, the disorder, or the interaction strength. Recent numerical studies\cite{Oganesyan2007,Pal2010,Monthus2010,Berkelbach2010,Chiara2006,Znidaric2008,Berkelbach2010,Bardarson2012,Canovi2011,Canovi2012,Luca2013,Kjall2014,Khatami2012,Iyer2013}, a renormalization group analysis \cite{Vosk2013,Vosk2014,Pekker2014}, as well as a rigorous proof subject to reasonable assumptions\cite{Imbrie2014} have strongly supported the existence of a many-body localized phase, at least in random spin chains. At the same time a useful phenomenological approach to describe the dynamics in the localized state has been developed\cite{Serbyn2013,Serbyn2013b,Huse2013a,Bauer2013,Swingle2013,Nanduri2014}.

The studies mentioned above, together with many others \cite{Yao2013,BarLev2014,Nandkishore2014b,Nandkishore2014a,Johri2014,Gopalakrishnan2014,Berkovits2014,Ros2014,Bauer2014}, have also shown that, far from being a boring dead state, the localized phase can display remarkably rich dynamical behavior.  In particular, it has been argued, that a number of distinct localized phases can exist\cite{Huse2013}, distinguished, for example, by presence of a broken symmetry or even quantum topological order in high energy eigenstates. In this case many-body localization protects quantum order, which could otherwise exist only in the ground state. Crucially, the distinct phases  have sharp signatures in measurable dynamical quantities\cite{Vosk2014,Bahri2013} and the system can exhibit universal singularities in the dynamics at critical points separating different localized phases\cite{Vosk2013,Vosk2014,Pekker2014}. It has also been demonstrated that quantum coherence of local degrees of freedom (i.e. at least single q-bit coherence) can be maintained for arbitrarily long times inside the localized phase\cite{Bahri2013,Serbyn2014}.

After reviewing the state of the art, and particularly the progress made in describing many-body localized states we shall turn to a discussion of some outstanding open questions. One such question is whether many-body localized states can serve as resources for quantum computation. Is it possible,
for example to utilize localization to protect quantum information encoded into entangled states of many q-bits? Another important open question concerns the nature of the many-body localization transition. The methods devised so far to address the localized phase rely, in one way or another, on the area-law, ground state like, entanglement entropy which prevails there. This simplifying feature ends at the critical point, beyond which the eigenstates are characterized by volume law entanglement. Furthermore,  if the transition is continuous, then we expect the localization length $\xi$ to diverge upon approaching the critical point. Because the area-law scaling of the entanglement is only established in sub-systems much larger than $\xi$ it becomes harder to compute the dynamical properties with methods that rely on the area law entanglement scaling as the system is brought closer to the critical point. Finally, we will touch on the important question how effects associated with many body localization and the localization transition can be observed in realistic systems.

\section{Many body localization and computability}\label{sec:MBL-Comp}

A good entry point for discussing the nature of many-body localized states is from the perspective of their computability.
In other words, can we achieve an efficient description of many-body localized states and their dynamics using a classical computer?
The backdrop to this discussion is the well known fact that a general quantum state is impossible to describe efficiently
because of the exponential growth of the  Hilbert space with the size of the system.

There are two well known exceptions to the exponential complexity of quantum problems, which are generic, that is, do not require fine tuning of parameters. 
 First, if one is interested
in thermodynamic properties or ground state correlations, then Quantum Monte Carlo provides a systematically controlled scheme for a large class of models. This is enabled by a mapping of the quantum partition function onto the partition function of classical models in one higher dimension.  Even when this mapping does not apply, efficient computation of ground state properties in one dimensional systems is enabled using the density matrix renormalization group (DMRG) method\cite{White1992} or, equivalently, using a matrix product state (MPS) ansatz\cite{Vidal2003,Perez-Garcia2007}. As we will discuss in more detail below, this scheme relies on the atypically small amount of quantum entanglement (area law entanglement entropy) present in ground states of local quantum Hamiltonians.  On the other hand, if we are interested in dynamics we must face the fact that the unitary time evolution generally allows the system to explore generic states in the Hilbert space that are characterized by extensive entanglement entropy and do not have an efficient description.
 In this section we will discuss how this problem can be partially overcome when the numerical calculation is done in the many-body localized phase. In later sections we will see how the same features which enable efficient computation, also facilitate the use of novel theoretical approaches to describe the dynamics in the many-body localized phase.

\subsection{Entanglement entropy}

We begin this section with a brief review of the notion of entanglement in many-body states and its quantification by the Von-Neuman entanglement entropy.
These concepts play a central role in modern many-body physics and are especially crucial for understanding many-body localization.
 To fully describe a generic quantum state of $N$ local spins, or q-bits, we need to specify $2^N$ complex numbers. The fact that our storage needs to grow exponentially with the system size is a consequence of the non-locality of quantum mechanics. Each of the $2^N$ coefficients cannot refer to a local section of the system, but rather to the weight of one global configuration of spins in the wave function.

We can of course write down states, which are completely local. These would be direct products of the quantum states of single spins
\be
\ket{\psi}=\prod_{i=1}^N\left(\sum_{\a=\ua,\da}c_{\a i}\ket{\a}_i\right)=\sum_{\{\a\}}\left(\prod_i c_{i\a} \right)\ket{\{\a\}}.
\label{product}
\ee
Clearly to specify this state we need only $N$ complex spinors $(c_{\ua i}, c_{\da i})$, which now define the state of local objects, the directions of the $N$ spinors on the Bloch spheres. Such an economical description is possible because there are no quantum correlations, which entangle different parts of the system. States that arise in physical situations can have different degrees of non-locality, varying between the non-entangled product states to the highly entangled generic states of the Hilbert space. To assess how much computational resources we need in order to describe a given state we need to quantify the degree of non locality. This will be done below using the notion of the entanglement entropy.

Specifically we ask how entangled is one part of the system $A$ with the supplementary part $B$ when the entire system is in a given quantum state. 
If the state is non-entangled then it is possible to write it as a product of states restricted to the two disjoint subsystems $\ket{\psi}{=}\ket{\psi_A}\otimes\ket{\psi_B}$. More generally, even if the state is not a product state, it is always possible to write it as a linear superposition of an orthogonal set of product states:
\be
\ket{\psi}=\sum c_n \ket{\f^A_n}\otimes\ket{\f^B_n}
\ee
where $\ket{\f^A_n}$ is an orthogonal  basis of the Hilbert space of the smaller subsystem $A$. This is called the Schmidt decomposition of the state $\ket{\psi}$. The reduced density matrix of the subsystem $A$ is easily computed in the Schmidt basis: $\rho_A= \tr_B\ket{\psi}\bra{\psi}=\sum_n |c_n|^2\ket{\f^A_n}\bra{\f^A_n}$.

The coefficients $c_n$ hold the non-local information on the correlations between the subsystem $A$ and the rest of the system. If more than one of them is non-vanishing then there is uncertainty as to what is the quantum state of subsystem $A$. The entanglement entropy $S_A= -\sum_n |c_n|^2 \ln |c_n|^2 =-\tr\rho_A \ln \rho_A$ quantifies this uncertainty or, from a different perspective, the complexity of the state $\ket{\psi}$. Indeed $e^{S_A}$ counts roughly how many states of the local Hilbert space of subsystem $A$ we need to keep in order to fully describe all measurements made there. Note that in the special case of a product state only one coefficient $c_1=1$ is non-vanishing and the entanglement entropy vanishes.

To characterize the entanglement structure and complexity of the state it is useful to examine how $S_A$ scales with the size of the sub-system $A$.
Consider a sub-system of volume $L^d$ embedded in a much larger (ideally infinite) system. The maximally entangled case is when all $2^{L^d}$ coefficients $c_n$ are of equal weight, leading to volume law entanglement $S_A\sim L^d$. In this case the quantum correlations between the two subsystems lead to complete uncertainty as to the state of the subsystem and the reduced density matrix is a unit matrix as in an infinite temperature state. It is interesting to note that if we pick a random state in the Hilbert space of the full system, then almost always, in fact with probability $1$ in the infinite system limit, it will behave as a maximally entangled state\cite{Page1993,Foong1994,Sen1996}. This gives an interesting perspective on thermalization. If the unitary time evolution eventually brings the system to a generic state, which it does if the dynamics is unconstrained to freely explore the Hilbert space, then the outcome would correspond to a maximum entanglement entropy state, just as in equilibrium.

As we already mentioned, quantum states, which follow a volume law entanglement scaling do not have an economical description because the information is held in exponentially many non-local coefficients.
Fortunately, there are states relevant to physical situations that are more tame. For example in all known cases, ground states of physical hamiltonians with only local interactions exhibit area law entanglement (i.e. $S_A\sim L^{d-1}$), possibly with a logarithmic correction. The logarithmic correction $S_A\sim L^{d-1} \ln L$ is found in certain gapless critical systems, e.g. conformal field theories in one dimension\cite{Holzhey1994,Vidal2003,Latorre2004,Calabrese2004,Calabrese2009} and free Fermions also in higher dimensions \cite{Gioev2006}  (see also \cite{Eisert2010} for a review). For gapped one dimensional systems the area law is proven rigorously \cite{Hastings2007}.

There is a simple intuitive picture of the area law entanglement when correlations are short ranged. In general every spin in sub-system $A$ that is entangled with a spin in the supplementary part $B$ contributes $O(1)$ to the entanglement entropy. Because correlations are short ranged, only pairs of spins residing on opposite sides of the boundary between A and B and sufficiently close to it can contribute to the entanglement entropy. Hence the entropy should scale as $\xi L^{d-1}$, where $\xi$ is the correlation length.

We will argue below that the localization length has a similar effect in many-body localized eigenstates.
Localization implies that quantum correlations between local objects (e.g. spins) have a finite extent set by the localization length. Thus we expect such states to follow similar entanglement scaling $S_A\sim \xi_{\text{loc}}L^{d-1}$.

\subsection{Matrix product states}

When applied to one dimensional systems, the area law entanglement of (non critical) ground states implies that their complexity does not increase with system size.
Indeed the low entanglement  of quantum ground states in one dimension is exactly what allows for efficient computation using the density matrix renormalization group (DMRG) scheme\cite{White1992,Schollwock2011}. The Hilbert space truncation scheme at the heart of the algorithm relies on the fact that only a few states are needed for the description of any sub-system embedded in the full system.
It took some time after the invention of the DMRG algorithm to realize that  it is in fact equivalent to a variational scheme which approximates the ground state of a one dimensional model within a family of states known as Matrix product states\cite{Ostlund1995,Dukelsky1998}.

Matrix product states (MPS) are direct generalizations of product states. Consider a chain of spin-like objects, each with $m$ internal states. As in the above example of the product state (\ref{product}), the starting point is a state of a single spin on site $i$: $\ket{\psi_i}=\sum_{\a=1}^m M^{(i)}(\a) \ket{\a}_i$. However, we now generalize the single-spin state by promoting the coefficients $M^{(i)}(\a)$ from  complex numbers to $D\times D$ matrices. Of course, now $\ket{\psi_i}$ is not a physical state of a single spin. But when we take the product of $\ket{\psi_i}$ over all the local spins then we obtain a valid {\em many-body} wave function when taking the trace over the resulting matrix products:
\be
\ket{\Psi}=\text{tr}\left[\prod_{i=1}^N\ket{\psi_i}\right]\equiv\sum_{\a_1,\ldots,\a_N=1}^m\text{tr}\left[M^{(1)}(\a_1)\cdot\cdot\cdot M^{(N)}(\a_N)\right]\ket{\a_1,\a_2,\ldots,\a_N}
\ee
The parameter $D$ is commonly known as the "bond dimension" and we note that when $D=1$ the MPS becomes a simple product state. On the other hand by increasing $D$ these wave functions can represent states with increasing degree of non locality, while the storage requirement is still only linear in the number of spins $~D^2 m N $.
It can be shown that a matrix product state has constant entanglement scaling with $S_A(L)\sim \log_2 {D }$.

So far we have discussed matrix product states as a computational scheme, used to obtain and characterize ground states of one dimensional models. However, there is good reason to expect that they can similarly provide an efficient description of many-body localized eigenstates at high energies in one dimension. Intuitively, it seems that localized  eigenstates should be required to support only area-law entanglement scaling, even if only to justify their name. This is because entanglement entropy growing faster with the sub-system size would imply presence of non-local quantum correlations in the state. Bauer and Nayak \cite{Bauer2013} gave further arguments to support area law entanglement, which rely on the definition of the many-body localized state as being perturbatively related to simple localized wave functions, such as the localized state of non-interacting Fermions. In section \ref{RG-eigenstates} we will  see that the constant entanglement scaling in generic one dimensional localized states obtains naturally from a renormalization group approach. We will also identify, as a bonus, special critical localized states which support $\log L$ entanglement scaling.
All of this is, of course, in sharp contrast to eigenstates in thermalizing systems, which by the eigenstate thermalization hypothesis are characterized by volume law entanglement.

There is an important difference however between the matrix product state description of ground states and that of MBL eigenstates. In the former there is an efficient algorithm that allows the MPS to converge to the correct ground state. In the case of high energy states there is no known way to target a particular state or energy. 

\subsection{Entanglement growth in time evolution}

Matrix product states can be used, at least in principle, to simulate unitary time evolution. This is implemented in the numerical scheme TEBD (time evolving block decimation)\cite{Vidal2003a}, which essentially solves a time dependent variational problem. The evolution operator is applied to an MPS for infinitesimal time-steps and the outcome is projected after each step back onto the family of matrix product states with the desired bond (matrix) dimension $D$. There is however a fundamental obstacle, which hinders running the scheme to long times. The problem is inescapable growth of the entanglement entropy in dynamics.

In a common setup for computing dynamics one is interested in the time evolution starting from a natural initial state with low entanglement. It could be, for example, a product state or a ground state of a simple initial Hamiltonian. But the buildup of correlations in the course of time evolution usually lead to linear growth of the entanglement entropy (of a sub-system) in time. In order to keep up with this entanglement growth within the MPS scheme it is necessary to increase the bond dimension $D$ exponentially in time, which rapidly becomes unfeasible.

\begin{figure}
\centering
\includegraphics[width=16cm]{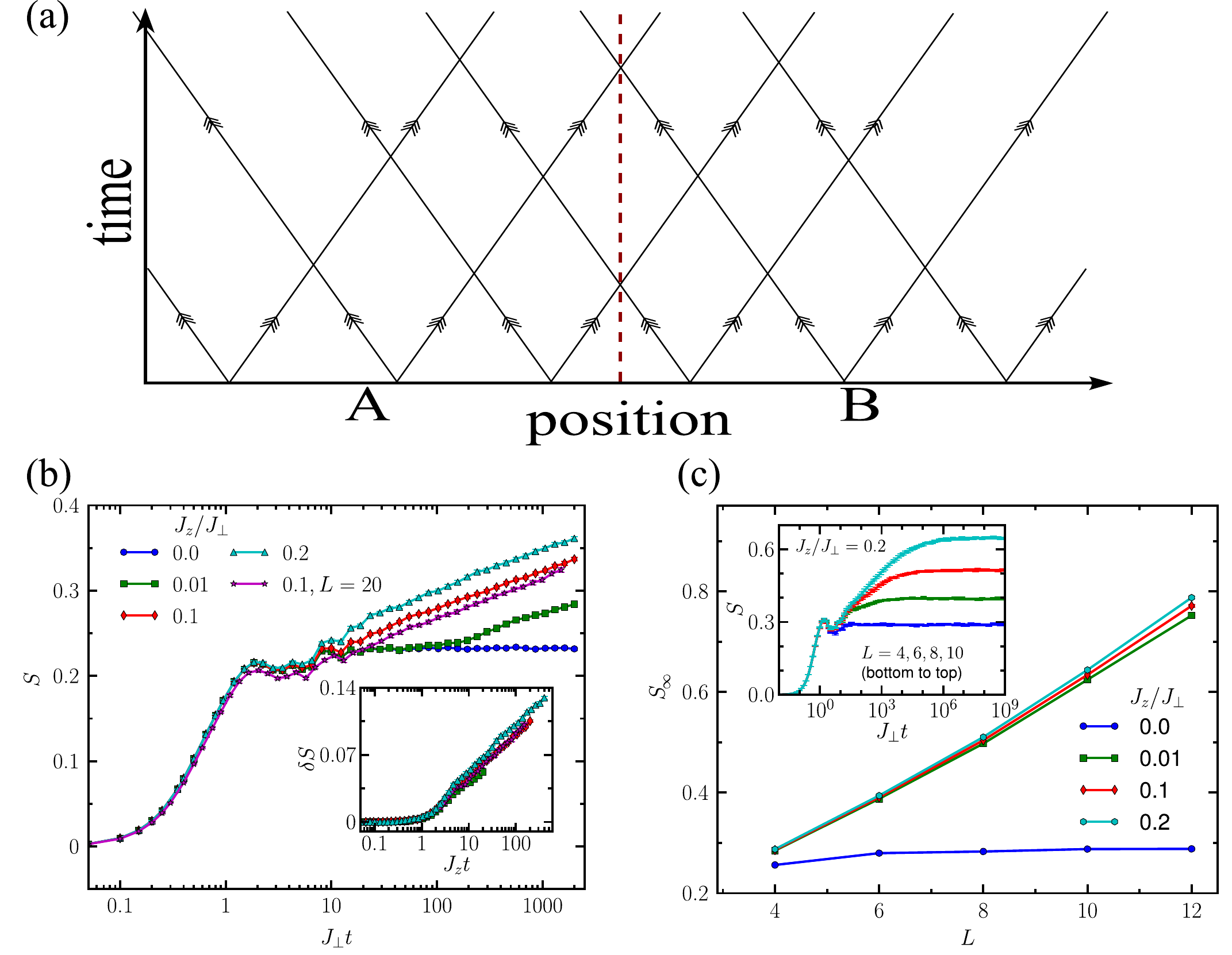}
\caption{ (a) Intuitive picture for $t$ linear entanglement growth in clean systems. Quasi-particles prepared in a localized initial state propagate as a superposition of a right and left moving particle, thus generating entanglement when the light-cone crosses the subsystem boundary. (b) Entanglement growth in a many-body localized system showing delayed logarithmic growth. Inset shows scaled plots with time measured in units of $1/J_z$, indicating that the delay is set by the interaction scale.  (c) Saturated value of the entropy  shows area law for the non-interacting system ($J_z=0$), and volume law for the system with interactions. The inset shows the time evolution up to the saturation value. Panels (b) and (c)  are reproduced from Ref. [\onlinecite{Bardarson2012}], copyright (2012) by The American Physical Society.}\label{fig:bardarson}
\end{figure}

There is a common intuitive picture for the linear growth of the entanglement entropy\cite{Calabrese2004} (see Fig. \ref{fig:bardarson}a). An initial non-entangled state can be viewed as one with a finite density of localized quasi-particle excitations. The quasiparticles begin to propagate with maximal velocity $\pm v$ to either left or right, thereby generating uncertainty, or $O(1)$ entanglement, each time such a light cone crosses the boundary of sub-system $A$.
Based on this intuition the entanglement would stop growing if quasiparticles were localized and MPS calculations may be carried out for long times.

Indeed TEBD simulations performed on random spin chains with strong disorder were able to reach quite long times, but the result was initially surprising. Instead of saturation, these studies found unbounded logarithmic growth of the entanglement entropy in time\cite{Chiara2006,Znidaric2008,Bardarson2012}. As an example, we show in Fig. \ref{fig:bardarson}(b), the result of a thorough numerical study by Bardarson et. al. \cite{Bardarson2012}. The TEBD calculations reported in this paper were done on the XXZ spin-$1/2$ chain with a random field:
\be
H=J_\perp\sum_i \left(S^x_i S^x_{i+1}+ S^y_i S^y_{i+1}\right)+J^z\sum_i S^z_i S^z_{i+1}+\sum h_i S^z_i
\label{Hxxz}
\ee
This Hamiltonian (\ref{Hz}) can be mapped, using a Jordan-Wigner transformation, to an interacting fermion model, where $J_\perp/2$ maps to fermion hopping, $J_z$ to the nearest neighbor interaction and $h_i$ is a random on-site potential.
The initial states for the dynamics were taken to be product states of spins randomly pointing up or down along $S^z$.

The growth of the entanglement entropy starting from the un-enetangled initial states is shown in Fig.  \ref{fig:bardarson}(b) for various values of the interaction $J^z$. For $J_z=0$, which maps to non interacting fermions, the entanglement entropy indeed saturates to a value independent of the system size. But for non-vanishing interaction $J^z$, the apparent saturation is replaced by onset of logarithmic growth of the entanglement, $S_A= S_0+s \log t$ after a delay time set by the inverse interaction energy.  These results suggest a fundamental difference between the dynamics of non-interacting Anderson-localized states and the dynamics in the many-body localized phase.
It is interesting to note that for non-interacting Anderson localized chains, $\log t$ growth of the entanglement entropy was proven to be a theoretical upper bound\cite{Burrell2007}.  But the results indicate that interactions are necessary to drive the entanglement growth and reach $\log t$ growth.

To check if  the growth of entanglement implies thermalization at long times, Bardarson et. al. computed the saturated value of the entanglement entropy of a finite sub-system of length $L$ using exact diagonalization. Because of the slow logarithmic growth of the entanglement, this saturation came only after a time exponentially long in $L$.  The saturation value of the entropy plotted as a function of $L$ in Fig. \ref{fig:bardarson}(c) shows a clear linear dependence, i.e. volume law entanglement.
However the entropy density in the system is significantly smaller than it would be in thermal equilibrium for the same energy density. The conclusion is therefore that the system does not attain true thermal equilibrium.

We have argued above that entanglement spreads ballistically in clean systems, whereas numerical evidence and theoretical arguments, to be discussed later, show that the entanglement spreads logarithmically in time in the many-body localized state.
A natural question then, is how entanglement spreads in delocalized random systems in which the energy spreads diffusively. Interestingly there is good evidence that the spread of entanglement, even in diffusive systems is nonetheless ballistic\cite{Kim2013}.

Why does entanglement propagate faster than energy? A heuristic argument can be given, which relies on the fact that entanglement is not a conserved quantity that needs to be transported from one side of the system to the other. To equilibrate a subsystem of size $L$ that has a lower energy density than the rest of the system,  $L$ quanta of energy, must be transported into it. But exchanging energy of order $1$ may be enough to fully entangle the sub-system\cite{Vosk-to-be-published}.

The numerical results on the entanglement growth raise a number of interesting questions. What is the (interaction driven) mechanism behind the seemingly universal growth of the entanglement as $log(t)$ in the many-body localized phase?  Are there other possible modes of entanglement growth in the localized phase? Can the difference between interacting MBL states and Anderson-localized non interacting particles be detected using physical local observables that would be amenable to experimental test?
What controls the  value of the entropy density reached in the long time steady state when it is not the thermal-equilibrium value? In the following sections the answers to some of these questions will be elucidated through the use of a systematic RG scheme as well as a simple phenomenological  description of the many-body localized phase.

\section{Renormalization group approach}\label{RG-eigenstates}

We now turn to discuss a renormalization group approach, which offers  insight on the nature of the many-body localized phase and allows controlled analytic as well as numerical calculation of various universal aspects of the dynamics.
As in our previous discussions, it will be useful to consider two complementary viewpoints of the localized phase, one which focuses on the time evolution and another which highlights the structure of eigenstates. We will see that such a scheme is ultimately enabled by the low entanglement entropy in the many-body localized phase, the same feature which allowed efficient numerical computation using matrix product states.

Standard RG schemes are aimed to capture the ground state and the asymptotic low energy excitations. In the process, high energy states are gradually integrated out, thereby renormalizing the effective interactions which govern the lower energies, while maintaining low energy correlations invariant. The hope is that the process converges to simpler fixed point theories. Such a truncation scheme cannot work for the time evolution problem, where we are specifically interested in the dynamics of high energy states. A revised RG philosophy is called for.

The goal in using RG to solve a time dependent problem at high energies is to obtain a universal description of the evolution on long time scales, while possibly losing information on the high frequency fluctuations. The RG scheme should provide a prescription how to successively integrate out  fast modes in a way that keeps the dynamics on longer time-scales invariant. The procedure is useful if the long time dynamics indeed possess robust universal features, captured by a simple fixed point theory. Fortunately this turns out to be the case in the Many body localized phase, at least in certain classes of random spin chains.

\subsection{RG scheme}
An RG approach for the dynamics of random spin chains at high energies was developed in references  \cite{Vosk2013,Vosk2014,Pekker2014}. The basic scheme is controlled by strong disorder, similar to the strong disorder
RG which captures the low energy physics of random spin chains \cite{Dasgupta1980,Fisher1992} (See [\onlinecite{Refael2013},\onlinecite{Igloi2005}] for recent reviews). There are however important differences, which stem from the need to capture long time dynamics at high energy density above the ground state.

The idea is nicely illustrated through application to a quantum Ising spin chain
\be
H=\sum_i  J^z_i  S^z_i S^z_{i+1}+h_iS^x_i + J^x_iS^x_iS^x_{i+1} .
\label{eq:ising}
\ee
where all the coupling constants are independent random variables.
We note that using a Jordan-Wigner transformation the Hamiltonian (\ref{eq:ising}) can be written as a fermion model with pairing terms. The coupling $J^x_i$ then represents an interaction between fermions, which makes the model generic. We shall assume that the other terms  in the Hamiltonian are typically much larger than $J^x$.

Consider first a system that is undergoing time evolution from a given initial pure state. The dynamics at the shortest time scales are fast oscillations, with frequency $\Omega$, of spins coupled by the largest terms in the chain: either a transverse field $h_i$ on a site or an Ising coupling $J^z_i$ on a link. If the disorder is strong, then other spins in the vicinity of the fast ones will typically be affected by much weaker couplings and therefore be essentially frozen on the timescale $\Omega^{-1}$.
The large separation in scales then allows a controlled  treatment of the dynamics of the slow spins in the presence of the fast ones using time dependent perturbation theory\cite{Vosk2013,Vosk2014}. The outcome is an effective Hamiltonian for the slow degrees of freedom, which no longer contains the fast scale $\Omega$. The operation is iterated, with a constantly decreasing cutoff $\W$. The result is a series of effective Hamiltonians with renormalized distributions of coupling constants, which describe the dynamics coarse grained to increasing time-scales.

A complementary perspective of the problem is gained from deriving the same effective Hamiltonian from {\em time-independent} perturbation theory performed on the many-body eigenstates\cite{Pekker2014}.  As in the standard ground state scheme \cite{Dasgupta1980,Fisher1992}, we separate the Hamiltonian into a fast and slow part $H=H_\W+V$, where $H_\Omega$ is the local Hamiltonian corresponding to the strongest  coupling on the chain (a bond or a site term as above). $H_\Omega$ splits the spectrum into nearly degenerate multiplets separated by the large energy $\Omega$ that are only weakly coupled to each other through $V$.
The RG step consists of a unitary Schrieffer-Wolf transformation applied in order to approximately decouple the high and low energy multiplets. This leads to the effective Hamiltonian
\be
H_{\text{eff}}=e^{iS}\left(H_\W+V\right)e^{-iS}-H_\W,
\ee
where  $S$ is obtained by solving for the requirement $[H_{\text{eff}}(S),H_\W]=0$ to first order in $V$. The transformation generates effective interactions acting within the subspaces that are second order in $V$ while ensuring decoupling between the sectors to this order (See Ref. [\onlinecite{Pekker2014}] for details). The operation is repeated to obtain a flow of effective Hamiltonians with decreasing cutoff frequency $\Omega$.

\begin{figure}
\centering
\includegraphics[width=15cm]{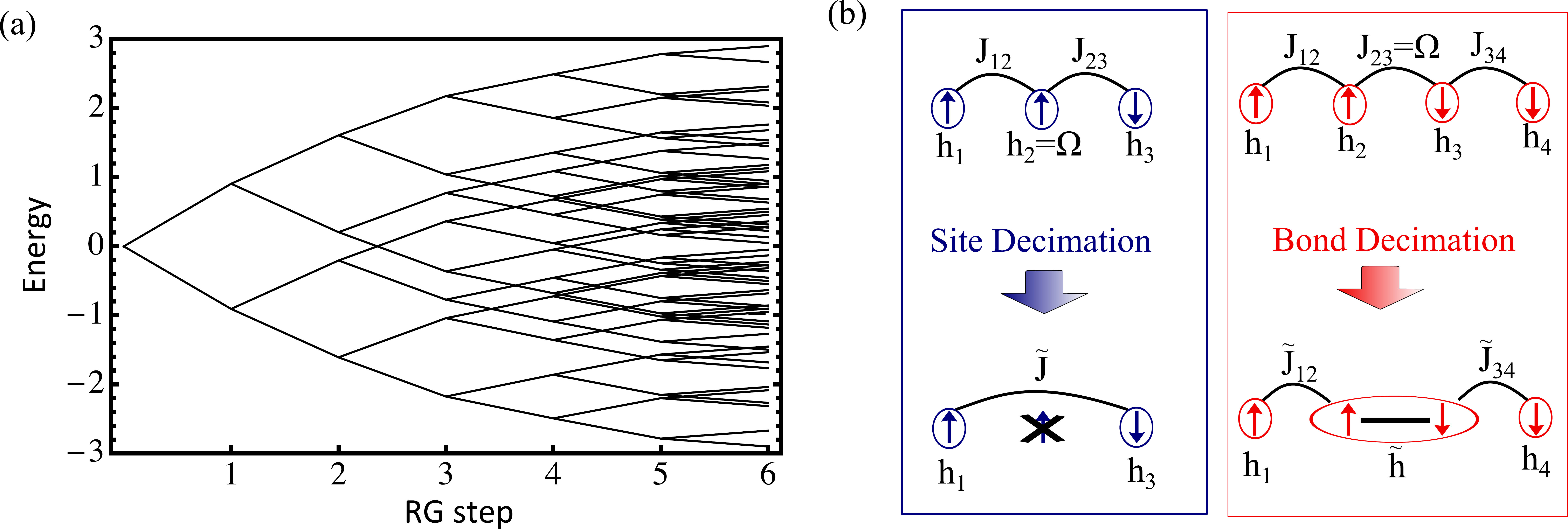}
\caption{ (a) \emph{Many body spectrum constructed itteretively in the course of renormalization.} The The leaves at the end of the branches correspond to the many-body energies (reproduced from Ref. [\onlinecite{Pekker2014}], copyright (2014) by The American Physical Society). (b) \emph{Illustration of the RG steps in an Ising chain \eqref{eq:ising}.} A large transverse field on a site freezes $S^x$ on that site and generates effective Ising coupling between the neighboring sites. Large Ising coupling on a bond fixes the alignment of the spins on the bond so that they act a single spin with effective transverse field.}\label{fig:rsrgx}
\end{figure}

The iterated RG process for constructing the many-body spectrum is illustrated in Fig. \ref{fig:rsrgx}a. Every RG step corresponds to branching of a tree into multiplets associated with the high and low energy subspace of the chosen strong cluster.  Thus with progression of the RG we resolve and focus on smaller energy {\em differences}, which correspond to ever lower frequency modes of dynamics.  Here lies the main difference between this scheme and the standard strong disorder RG \cite{Dasgupta1980,Fisher1992}, whereby we would have to freeze the strong cluster at every step to the {\em low energy} manifold. Thus the ground state scheme targets low absolute energies, while the dynamical scheme focuses on small energy differences, which occur in the entire spectrum.

For concreteness we demonstrate the RG scheme as applied to the quantum Ising model (\ref{eq:ising}).
First, consider the situation where the largest coupling is a transverse field $|h_i|=\Omega$ on a site. The effective Hamiltonian which governs the dynamics in the immediate vicinity of the chosen site is
\be
H_{\text{eff}}={2J^z_L J^z_R\over h_i} {\tilde S}^z_L {\tilde S}^x_i {\tilde S}^z_R+\sum_{\a=R,L}(J^x_\a {\tilde S}^x_i +h_\a){\tilde S}^x_\a,
\label{Heffh}
\ee
where $R,L$ denote the spins to the right and left of the site $i$.
Note that the renormalized operator on the fast site ${\tilde S}^x_i =e^{iS}S^{x}_i e^{-iS}$, is now an integral of motion. It commutes with all the terms of the effective Hamiltonian. The existence of such an integral of motion and its interaction with other integrals of motion is an important property of the localized phase, which we will discuss shortly. However, for the purpose of subsequent renormalization steps we can fix ${\tilde S}^x_i$ randomly to one of its eigenvalues $\eta=\pm\half$ and eliminate the site $i$ as an active site on the chain. Then the effective Hamiltonian (\ref{Heffh}) assumes the same form as the original Hamiltonian, with renormalized coupling constants: $J^z_{ren}=2 \eta J^z_L J^z_R/h_i$, $h^{ren}_{\a}=h_\a+\eta J^x_\a$.

Another type of RG step is required when the largest coupling is an Ising interaction on a bond $|J^z_i|=\Omega$. Then, the two spins are combined to an effective single spin $S^\a_n$, acting in a two dimensional Hilbert space of either the high or low energy subspace. e.g. for $J^z_i<0$ the two states in the low energy sector are the aligned states $\ket{\ua\ua}$, $\ket{\da\da}$ and in the high energy sector they are the anti aligned states. Again the RG transformation creates a conserved operator $\hat{\eta}=4S^z_i S^z_{i+1}$ which represents the relative spin alignment on the joined sites. As before, treating this operator as a number $\eta=\pm 1$ for the sake of subsequent renormalization leads to an effective Hamiltonian with renormalized values of coupling constants. For example, the new joined spin is subject to an effective transverse field $h_{ren}=J^x_i+ \eta h_i h_{i+1}/ \Omega$. Both of the RG decimation steps are illustrated in Fig. \ref{fig:rsrgx}b.

The flow of the distribution of coupling constants induced by the sequence of RG transformations can be analyzed by exactly the same methods that have been developed for the standard ground state SDRG scheme\cite{Dasgupta1980,Fisher1992,Igloi2005,Refael2013}.
However as we discuss in the following sections the interpretation of the results and the way physical properties are extracted from the flow can be very different.
On the one hand the RG scheme can be used to compute the instantaneous expectation values of observables in a system undergoing time evolution\cite{Vosk2013,Vosk2014}. On the other hand the scheme has been used as a way to hierarchically construct  the many-body spectrum and the eigenstates in localized systems  \cite{Pekker2014}. The latter scheme, termed RSRG-X (i.e. real space RG for excited states), can be used to compute low frequency response and correlation functions of localized systems at arbitrary temperatures. The idea is to construct only a small fraction of the tree illustrated in Fig. \ref{fig:rsrgx}a, where the branches are sampled by a Monte-Carlo Metropolis algorithm. A branch is accepted or rejected according to the thermal weight implied by the energy found at its end. In Ref. [\onlinecite{Pekker2014}] the frequency dependent thermal conductivity as well as spin auto-correlation functions were computed for the transverse field Ising model, showing universal singularities associated with a dynamical phase transition at high temperature (see discussion in section \ref{sec:qcp}).

\subsection{Limitation of the RG scheme -- resonances}

There is an important difference between the dynamical RG scheme discussed above and the standard ground state RG scheme  in the way the scheme can fail. As we shall see below the dynamical scheme requires the system to be in the localized state. It is invalidated by the same processes, which would lead to delocalization.

Consider an RG step in which we eliminate a site with a strong transverse field $|h_i|=\Omega$.
In a subsequent  RG step we may eliminate another spin with only slightly smaller field $|h_j|=\Omega-\delta\Omega$ located at a remote site $j$ on the chain. The alignment of each of those spins encoded by the eigenvalues of $S^x_i$ and $S^x_j$ are transformed by the RG steps into conserved quantities because locally flipping the spins would correspond to a high frequency mode. This however, neglects the possibility of an interaction that exchanges the $S^x$ orientation between the two sites $J_{\text{eff}} (S^z_i S^z_j+S^y_iS^y_j)$, which would only cost the low energy difference $\d\W$. Such an interaction, may be induced at a stage when all spins between the two sites $i$ and $j$ had been eliminated. The interaction leads to a low frequency mode delocalized between the two sites only if it is resonant, that is, if $J_{\text{eff}}\gtrsim \d \W$.
Such a resonant exchange cannot not occur in the ground state scheme, where the spins in $i$ and $j$ are both frozen to their low energy manifold.

In addition to the two site resonances, we may have long range resonances involving any number of sites.
The RG scheme can remain tractable and local only if such resonances do not proliferate, which is synonymous with the eigenstates being localized. The probability of two site resonances has been analyzed within the framework of the dynamical RG and it was shown that for sufficiently strong disorder their density on the chain vanishes\cite{Vosk2013}.
A more systematic analysis of resonances including any number of sites has been carried out as part of a rigorous proof of the stability of the localized phase in quantum spin chains\cite{Imbrie2014}.

\subsection{From RG to entanglement entropy}\label{sec:RG-ent}

The dynamical RG scheme can be used to characterize two different aspects of entanglement in the many-body localized phase: entanglement entropy of high energy eigenstates and growth of entanglement in time evolution. The two calculations require to adopt two complementary perspectives or interpretations of the RG flow.

\begin{figure}
\centering
\includegraphics[width=15cm]{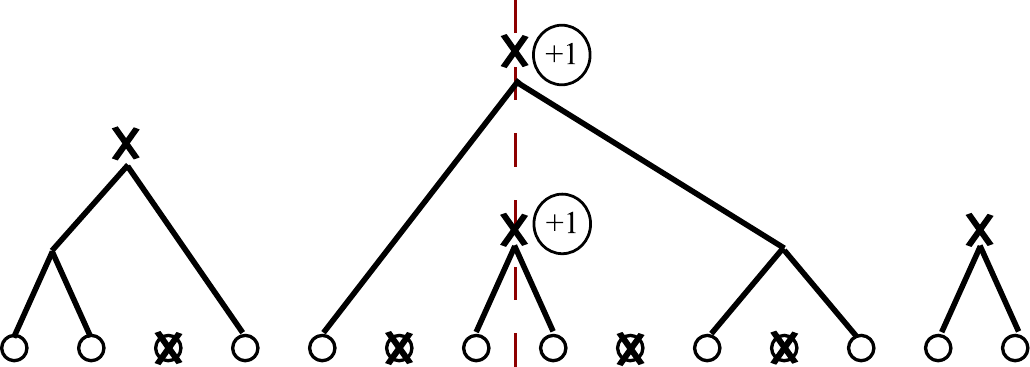}
\caption{{\em Entanglement entropy of an eigenstate as calculated in the RG scheme.} A tree structure in real space is obtained due to joining of sites into larger clusters in bond decimation RG steps. A cluster ends with an $X$ when it is decimated by a large transverse field. Every eliminated cluster which extends across the boundary between the two subsystems adds one unit to the entanglement entropy of the eigenstate. The tree like structure ensures that entanglement entropy does not grow faster than $\log L$.}
\label{fig:rgtree}
\end{figure}

{\bf Entanglement entropy in eigenstates --} To characterize the entanglement entropy in eigenstates we take the viewpoint of the RG as an iterative scheme to construct eigenstates. Because the integrals of motion obtained in each RG step take a fixed value in eigenstates we can
treat them as numbers that depend on  whether we take the upper or lower branch at each step. Thus, apart from the fact that we take a random branch instead of the lower branch, the scheme for constructing a high energy eigenstate is the same as for constructing the ground state. Therefore, the entanglement entropy can be computed in essentially the same way as done for the ground state using the SDRG \cite{Refael2004,Refael2007}.  In the case of the Ising model for example, a unit of entanglement is added on every step where a cluster of original spins that extends across the interface is eliminated due to a large (effective) transverse field on it. This calculation gives ground state like entanglement with scaling of $S\sim \log L$ of the sub-system size $L$ obtained at a critical point separating two distinct localized once dimensional phases\cite{Huang2014}.

The fact that the entanglement entropy cannot scale faster than  $\log L$ with sub-system size is due to the tree like structures in real space which are created in the course of renormalization. As depicted in   Fig. \ref{fig:rgtree}, a local tree is built when pairs of sites, or clusters, are joined into bigger clusters. Each local tree ends at the level in which the corresponding cluster becomes effectively frozen due to a large transverse field acting on it. In this scheme, there is no entanglement between trees representing  distinct clusters. But such entanglement would be generated if there were long range resonances, as discussed above, between the low and high energy subspaces of remote clusters.  Proliferation of such resonances would destroy the local structure of the tree, which in turn, would allow emergence of volume-law entanglement entropy.

{\bf Entanglement growth in time --} To understand how entanglement is generated in time it is better to think of the RG steps as a scale by scale solution of the time evolution problem. Consider for example a cluster that is eliminated due to a large transverse field on it. 
As before, the effective spin $S^x_i$ of this cluster is transformed to a conserved quantity, however in contrast to the eigenstate computation we cannot simply replace it by one of its eigenvalues. Because the initial state fixes the local spin projection $S^z_i$, rather than $S^x_i$, this degree of freedom remains dynamic and undergoes oscillations at a frequency of the order of the running cutoff $\W$.  In addition the second term in Eq. (\ref{Heffh}) generates interactions of the form $S^x_i S^x_j$ between the effective spins representing disjoint clusters.

There are now two sources of entanglement growth. First, every eliminated cluster, which extends across the interface, adds one unit to the entanglement entropy. This is essentially the same as the entanglement computed in an eigenstate and therefore the total entropy generated by this process at long times  is bounded by $\log L$. The more important source of entanglement is the dephasing of oscillations on different clusters due to the interaction between them which eventually leads to volume law entanglement $S\sim L$. The growth of the entanglement in time due to this process scales as $\log t$ in generic localized states, while it follows an enhanced growth as  $(\log t)^\a$ with $\a>1$ a universal exponent at certain dynamical critical points \cite{Vosk2013,Vosk2014}. Below we will give a simpler Heuristic understanding of the $\log t$ entanglement growth obtained for generic states.

\subsection{Integrals of motion}\label{subsec:IOM}

One of the results of the renormalization group approach is the emergence of local integrals of motion, that are gradually constructed in the course of the flow. Ultimately, the fixed point of the flow is an integrable Hamiltonian, which can be written in terms of these conserved operators. In this sense, many-body localized phases, can be understood in direct analogy to zero temperature quantum phases. These too can be viewed as stable RG fixed points and likewise can often be written in terms of emergent integrals of motion. For example a Fermi liquid is a quantum phase, with a fixed-point hamiltonian written in terms of the conserved  momentum-space densities $n_k$.

Note however, that because conventional RG targets low energies, the emergent conserved quantities are, correspondingly only operators which relate to asymptotically low energy excitations. For example, in Fermi liquids only physical fermion densities at momentum $k\to k_F$ have a finite overlap with the integrals of motion, a set of measure zero in the entire many-body Hilbert space. By contrast when the entire many-body spectrum is localized, the dynamical RG reveals a set of integrals of motion that spans the entire Hilbert space. Indeed, the absence of thermalization is attributed to the strong constraints imposed by these integrals of motion on the dynamics.

{\bf Phenomenological theory --}  It is worth pursing the analogy with quantum phases further. A particularly appealing feature to consider is that the RG fixed points which describe certain quantum phases also provide the basis for a useful phenomenological description. A case in point is Fermi liquid theory. At the fundamental level, it is good to know that it emerges as an RG fixed point, but it was originally conceived and is most commonly used as a phenomenological theory. In similar spirit, a useful phenomenological description of many body localized states has been developed\cite{Serbyn2013,Huse2013a,Serbyn2013a}. This phenomenological approach is described in some detail in the review by Huse and Nadkishore \cite{Nandkishore2014}.

Let us summarize the basic idea by considering the Hamiltonian
\be
H(\lambda)=H_0+\lambda H_z= \sum_i h_i \s^x_i +\sum_i V_i \s^x_i \s^x_{i+1}+\lambda\sum_i J^z_i\s^z_i\s^z_{i+1}
\ee
For $\lambda=0$ all local operators $\s^x_i$ commute with $H_0$ and it is trivially in the localized phase. For sufficiently small $\lambda$ the entire spectrum is expected to remain localized. It was argued in Ref. \cite{Serbyn2013,Huse2013a,Serbyn2013b} that in this case the set of bare operators $\s^x_i$ must map continuously to a new set of mutually commuting dressed integrals of motion
$
{\tilde \s}^x_i (\lambda)=Z_i(\lambda) \s^x_i + \text{tail},
$
which transform in the same way as the bare $\s^x_i$ operators under the global symmetries of $H$. The renormalized integrals of motion are quasi-local in the sense that they have a finite overlap with a strictly local operator. But they are generally dressed by a tail consisting of non local operators, such as $\s^x_l \s^x_j$ or $\s^z_l \s^z_k \s^x_k$, with coefficients that decay exponentially with the range. The operator overlaps $Z_i$ between strictly local operators and the true integrals of motion will vanish upon crossing the transition to the delocalized phase.

The unitary operator which transforms the bare $\s^x_i$ operators to the renormalized ones also brings the Hamiltonian into a classical diagonal form written entirely in terms of the $\ts^x_i$.
\be
H_{eff}= \sum_i \tilde{h}_i \ts^x_i+\sum_{i,j} \sum_n\tilde{J}^{(n)}_{ij}\,\ts^x_i \,\hat{B}^{(n)}_{ij}\,\ts^x_j
\label{Heff}
\ee
Here the operators $\hat{B}^{(n)}_{ij}$ represent all possible products of $\ts^x_l$ and $\ts^0_l$ (unit operators) over the sites $k$ between $i$ and $j$ ($i<k<j$).
The coefficients of the non-local terms $J_{ij}^{(n)}$ decay exponentially with the range $|i-j|$ as does the total effective interaction between a pair of spins at a given range, that is, $J_{ij}^{\text{eff}}=\sum_n \tilde{J}^{(n)}_{ij} \sim  V e^{-|i-j|/\xi}$, where
$V$ is the bare interaction and $\xi$ is the localization length.

Using the Hamiltonian (\ref{Heff}) it is easy to gain a simple intuitive picture\cite{Huse2013a,Serbyn2013} of the logarithmic growth of the entanglement entropy in the localized phase, seen in numerical simulations \cite{Chiara2006,Znidaric2008,Bardarson2012} (see  Fig. \ref{fig:bardarson}). Consider an initial state in which the dressed spins are not in eigenstates of $\ts^x$. For instance they may be pointing roughly along $\ts^z$. To compute the entanglement entropy consider a pair of such spins, which reside on two sides of the boundary between subsystem $A$ and the supplementary subsystem $B$. The two spins, separated by a distance $l=|i-j|$ are inevitably entangled with each other after a time $t\sim  1/J^{\text{eff}}_{ij}\sim V^{-1} e^{l/\xi}$ due to the effective interaction between them. By this time however all the spins within the interval $l$ between these two spins will have been entangled with each other through the larger effective interactions that act between them. Therefore the total entanglement entropy generated by this time scales as $S_A\sim l$. Inverting the above relation between $t$ and $l$ we get $S_A(t)\sim \xi \log_2 (V t)$.

As emphasized above, the Hamiltonian (\ref{Heff}) can be viewed as an RG fixed point. However it is also useful as a purely phenomenological description
of the dynamics deep in the localized phase, independent of the RG scheme that could have been used to derive it. In section \ref{sec:qi} below we shall see how the phenomenological approach can be applied to describe dynamical measurements aimed at characterizing the quantum coherence present in the localized phase. At the same time, the phenomenological approach is not sufficiently powerful to describe the universal critical dynamics which obtains at critical points separating distinct localized phases. Application of the RG scheme to describe such universal critical dynamics will be the topic of section \ref{sec:qcp}.

Finally, we note that the mutual eigenstates of all the integrals of motion $\ts^x_i$ defined above span the entire Hilbert space. Hence the Hamiltonian (\ref{Heff})  describes a situation in which all eigenstates are many-body localized. It is an interesting open question whether one can define local integrals of motion, which span part of the Hilbert space when the many-body spectrum has a mobility edge.

\section{Dynamical quantum phases and phase transitions}\label{sec:qcp}

An interesting observation, first made by Huse et. al. \cite{Huse2013}, is  that many-body localized eigenstates can belong to distinct classes characterized by different types of order. Because such eigenstates are not constrained by the ETH, the order they possess may be different than the order that would obtain had the system been coupled to a thermal bath. Indeed, the ground state like entanglement structure of many-body localized eigenstates suggests that even at high energy density they can sustain order that could otherwise be found only at zero temperature. For example, a one dimensional random Ising spin chain (\ref{eq:ising}) can have high energy eigenstates characterized by broken $Z_2$ symmetry, which would be forbidden in a one dimensional system at finite temperature. In the same way many-body localization can protect topological order\cite{Huse2013} as well as topological edge states, allowing them to exist even at high energies\cite{Bahri2013,Chandran2014}.

The distinct localized phases correspond to RG fixed points characterized by very different sets of integrals of motion. For example, in the paramagnetic localized state of the quantum Ising spin-chain (\ref{eq:ising}) the integrals of motion are related to the local spin operators $S^x_i$ and are even under the $Z_2$ symmetry, whereas in the broken symmetry phase the integrals of motion are directly related to the $S^z_i$ and are odd under the symmetry.

The dynamical RG approach discussed above is ideally suited to describe phase transitions between different many-body localized states\cite{Vosk2014,Pekker2014}, where the integrals of motion change their nature. In the Ising model, for example, the transition to the broken symmetry phase is described as successive joining of spins to an infinite static cluster with random $S^z$ orientations. This situation describes broken symmetry in each eigenstate. But since the spin orientations are different in different eigenstates there is no spin expectation value in a thermal ensemble. Hence the signatures of
the transition can only be seen in dynamical and not in thermodynamic measurements.

\noindent {\bf Critical point in the time evolution --}
Consider the time evolution of the extended Ising spin chain  (\ref{eq:ising}) prepared in a simple product state with spins randomly oriented along $\pm\hat{z}$. We may ask how the local expectation values $\av{S^z_i(t)}$ decay with in time. In the broken symmetry phase there is a non-vanishing probability that site $i$ is part of the infinite static cluster, in which case the above expectation value saturates to a non-vanishing value.
It is tempting to take the disorder average over this saturated value to be the order parameter. But this quantity in fact vanishes   because the saturation value may be positive as well as negative in different realizations of disorder. Instead we should take
$\Psi\equiv \overline{\av{S^z_i(\infty)}^2}$, which is a dynamical version of the Edwards-Anderson (EA) order parameter for spin glasses\cite{Edwards1975}. Solution of the RG flow shows that the dynamical order parameter onsets as $\Psi\sim |\D|^{2-\phi}$  \cite{Vosk2014}, the same as the ground state order parameter, when the system is tuned across the quantum phase transition.

%
As mentioned above, the two phases are described by different sets of integrals of quasi-local integrals of motion. At the critical point, however, a fraction of the integrals of motion become somewhat non-local with the non-locality encompassing all scales. This has interesting consequences on the entanglement growth, which can also be calculated using the RG scheme.
The solution of the RG flow can be used to compute the entanglement growth in time. As explained in  section \ref{sec:RG-ent} this comes from the induced interaction $J^x_{\text{eff}}$ between the conserved operators  $\tilde{S}^x_i$.  It was shown in Ref.
\cite{Vosk2014} that at the critical point this leads to  $S(t)\sim \Theta( t- \hbar / J^x_{0})[\ln t]^{2/\phi}$, where $2/\phi\approx 1.236$ is a universal exponent. That is, the entropy growth  is enhanced at the critical point compared to the generic  $\ln t$ behavior.

\noindent {\bf Finite temperature dynamical phase transitions --}
We now discuss, following Ref. \cite{Pekker2014}, a different setup for a dynamical experiment. Consider again a system described by the quantum Ising model (\ref{eq:ising}), but now weakly coupled to a thermal (e.g. phonon) bath with temperature $T$. This puts the system into a mixed state described by a thermal canonical ensemble, which of course, cannot sustain the localization protected quantum orders discussed above.
Can such a phase nonetheless exhibit universal dynamical phenomena, which reveal the localized nature of the underlying eigenstates and, in particular, show sharp universal signatures of the eigenstate transitions?
%

To answer this question first assume that the coupling to the bath is infinitesimal, so that it serves only to prepare the thermal ensemble but does not lead to relaxation at finite times. Under these conditions, consider
the spin auto-correlation function for a specific disorder realization $\av{S^z_i(t) S^z_i(0)}_T=\tr\left[\rho(T) S^z_i(t) S^z_i(0)\right]$. Depending on system parameters, and possibly the temperature, the eigenstates making up the ensemble and thus contributing to this correlation function may be either in the paramagnetic or the glass state. The autocorrelation function taken in glass eigenstates saturates to a positive value $m(E)^2$ at long times  and it therefore does so also in the thermal ensemble  $\av{S^z_i(\infty) S^z_i(0) }_T=m_i(T)^2$ when the system is in the glass phase. The disorder average of this quantity $\Psi= \overline{m_i(T)^2}$ is an Edwards-Anderson (EA) order parameter of the glass phase. Thus, the dynamical measurement of  spin autocorrelations offers a sharp signature of the underlying eigenstate transition even when the system is in the thermal ensemble.

In solids, where the coupling $\eta$ to a phonon bath is non-vanishing, the spin autocorrelation function would decay on a relaxation time $\tau=\eta^{-1}$. Then, the saturation value of the auto-correlation function appears only at intermediate times $<\tau$ and the transition broadens to a universal crossover.
There is a clear analogy here with standard quantum critical points (QCPs). While QCPs are strictly defined only at $T=0$, they have a much wider sphere of influence giving rise to universal crossover phenomena in which temperature sets the only scale\cite{Sachdev-book}. Similarly the finite temperature dynamical quantum-critical points in the localized phase are sharply defined only when the bath coupling $\eta$ is set to zero. But they control universal crossover phenomena, which may be seen at finite coupling to the bath, whereby this coupling sets the only scale.



The dynamical RG method offers a powerful approach to study the universal crossovers associated with the finite temperature dynamical quantum critical points.
Specifically, the method allows to compute the low frequency limit of correlation functions at arbitrary temperature \cite{Pekker2014}. This is in contrast with the standard equilibrium scheme which can find dynamic correlations only at zero temperature (or in the limit $T\ll \w$) \cite{Motrunich2001}. Thermal expectation values are calculated through Monte-Carlo sampling of the tree shown in Fig \ref{fig:rsrgx}(a) as described in section \ref{RG-eigenstates}.
The method was used to obtain the phase diagram of the extended Ising model (\ref{eq:ising}), universal finite size scaling functions of the EA order parameter, as well as dynamic response functions.

\section{Quantum information and coherence in MBL states}\label{sec:qi}

In this section we consider the dynamics of many-body localized states from the point of view of  quantum information. We do this with two goals in mind. First, we may ask how such localized media can be useful for protecting quantum information from decoherence. Second we can use measurements of q-bit coherence as sharp diagnostics to characterize the nature of the localized state and discern between different phases.

We have seen that many-body localized states are characterized by a large set of (quasi) local integrals of motion. It is tempting to utilize these operators as q-bits in order to store quantum information. But there are at least two major difficulties in applying this program, which have to do with the nature of the dynamics in the localized state.
For simplicity let us assume that the integrals of motion can be written as spin-$\half$ operators, say $\ttau^z_i$. The first apparent problem in using these objects as q-bits is that if we initialize the q-bit in some arbitrary direction on the Bloch sphere, only the projection on the $z$ axis is conserved. The orthogonal directions will precess with undetermined frequency, which depends on the interaction with the other q-bits and therefore may be hard to track. Indeed the logarithmic growth of the entanglement entropy is evidence that some kind of slow decoherence is taking place. The second problem is that the intended q-bits are not strictly local operators. In the localized phase they have an overlap with the constituent degrees of freedom $\tau^z_i$, but they also include tails, albeit exponentially decaying, made of  multiple spin operators. It is therefore not clear how to access the dressed spins $\ttau^\a_i$ in order to extract their coherence.

In what follows, we describe measurements and manipulations performed directly on the bare constituent degrees of freedom (i.e. $\tau^\a_i$ ), which allow to quantify the quantum coherence left in them after some time. We explain how such measurements can be used to diagnose the many-body localized phase and briefly discuss the potential for quantum information applications.

\subsection{Spin echo}
Ramsey type experiments have been proposed as diagnostics of the local coherence present in different classes of many-body localized states\cite{Bahri2013,Serbyn2014}. Following Ref. [\onlinecite{Serbyn2014}] we review such a protocol for the simplest MBL state in a spin-$\half$ system. In this case the effective hamiltonian can be written, as described in section \ref{subsec:IOM}, using the complete set of quasi-local integrals of motion
\be
H_{eff}= \sum_i \tilde{h}_i \ttau^z_i+\sum_{i,j} \sum_n\tilde{J}_{ij}\,\ttau^z_i \ttau^z_j + \ldots
\label{Hz}
\ee
As a natural starting point we assume the system is prepared in a thermal ensemble, as could be done for example by coupling to a bath, then decoupling once the system had thermalized.
We want to use the eigenstates $\ket{\tilde{\ua}},\ket{\tilde{\da}}$ of one integral of motion $\ttau^z_i$ as a test q-bit. For now we pretend that these states can be  addressed and manipulated directly in experiment as if $\ttau^\a_i$ were strictly local operators. The q-bit is initialized in a pure state $\ket{\tilde{\ua}}+\ket{\tilde{\da}}$. This can be done, for example, by measuring it in the $\ttau^z_i$ basis and then applying the appropriate rotation around the $\ttau^y_i$ axis. The question we ask is whether we can retrieve the initial state after it had undergone arbitrary long time evolution subject to the effective Hamiltonian (\ref{Hz}).

In the way we set up the problem, the answer is clearly positive.
In each state of the initial ensemble, all spins other then the test q-bit remain completely static in their $\ttau^z_j$ eigenstates. The test q-bit precesses in the static field that the other spins collectively impart through the effective interactions with it. Therefore, a unitary $\pi$ rotation of the test q-bit about the $\ttau^x_i$ axis
applied at time $t_R$ would perfectly reverse the precession and revive the initial state at time $2t_R$.

The procedure described above relied on the assumption that we can directly access and manipulate the states of the dressed operators $\ttau^z_i$. In reality, however, we can only access and implement unitary operations on the states of the strictly local physical spin $\tau^z_i$. Nonetheless it was demonstrated that a partial echo can survive to infinite time \cite{Bahri2013,Serbyn2014}. To compute the dynamics in this case we have to separate it to external operations, such as $\pi/2$ and $\pi$ pulses, which are easily computed in the basis of physical spins, and the intrinsic hamiltonian evolution, which is simple in the dressed basis:
\be
\ket{\psi(2t)}\approx e^{i {\pi\over 4}{\tau^y_i} }e^{-ih_{\text{eff}}\ttau^z_i t}e^{-i {\pi\over 2}{\tau^x_i}}e^{-ih_{\text{eff}}\ttau^z_i t}e^{-i {\pi\over 4}{\tau^y_i}}\ket{\psi(0)}
\ee
Each basis change made to match between seguents of the evolution leads to suppression of the echo by a Frank-Condon factor proportional to the overlap between the  local and dressed basis~\cite{Bahri2013}
\be
|\bra{\tau_i}\ttau_i\rangle|^2\sim e^{-a\xi}.
\ee
Hence the finite localization length protects the spin echo from being suppressed by a complete orthogonality catastrophe.

Together, the conservation of the $z$ component of the spin and the retrievable transverse component, provide a sharp signature of the many-body localized phase. The first effect indicates the presence of a local conserved quantity and the second demonstrates the persistence of local quantum coherence in the state. {\bf These effects allow to distinguish between a quantum many-body localized state and a classical glass.}

Relaxation processes in a classical glass can be extremely slow
showing, for example, logarithmic in time relaxation of certain observables\cite{Amir2011}. In theory this is distinct from the MBL state, where local observables with non-vanishing overlap with the integrals of motion do not relax at all. But in practice it may be hard to discern ultra-slow relaxation from complete absence thereof. On the other hand, while classical observables relax slowly in a classical glass, quantum coherence decays rapidly.
For this reason a spin echo experiment, which directly measures the persistence of quantum coherence in local degrees of freedom can serve as a much more sensitive probe of the MBL state.

{\bf Distinction between many-body localization and non interacting Anderson localization --}
We argued in section \ref{subsec:IOM} that the unbounded logarithmic growth of the entanglement entropy in the time evolution starting from a non-entangled state can be understood as a result of interaction induced dephasing, which does not take place in a non-interacting Anderson localized state.
It is natural to ask if this distinction between the interacting and non interacting states can be seen in spin-echo experiments, which directly measure phase coherence.

A spin-1/2 chain model is said to be "non-interacting" if the system can be mapped through a Jordan-Wigner transformation to a quadratic Fermion model. In this case the effective Hamiltonian (\ref{Hz}) written in terms of the dressed spins does not include extended interactions such as the $J_{ij}$. The simple spin echo protocol discussed above does not distinguish between the non-interacting and interacting cases.
However Serbyn et. al. \cite{Serbyn2014} proposed a modified protocol, which does discern the two cases.
The scheme goes through just like the simple echo protocol up to and including the $\pi$ pulse applied to reverse the precession of the test spin. The new element of the scheme is that after the $\pi$ pulse one also rotates $N$ spins in a region far away from the test spin. If there are interactions, then this change would lead to mismatch between the evolution of the test spin before and after the $\pi$ pulse and thereby to suppression of the echo. On the other hand in the non-interacting Anderson localized state the echo of the test spin would not be sensitive to manipulation of far away spins.

{\bf From a single q-bit to many --}
So far we discussed an echo protocol for a single test spin as a diagnostic of the localization properties. An interesting open question with regard to quantum information applications is whether the localized state can similarly help preserve the quantum information stored in a many q-bit entangled state. A direct extension of the single spin echo protocol, is not practical as it would require a cycle of $2^N$ operations to preserve a general state of $N$ spins.

A possible approach for storing quantum information in a many-body localized medium would be to use only a small fraction of the spins in the MBL state as the q-bits. If the distance $l$ between the designated q-bit spins is much larger than the localization length then the interactions between the q-bits is suppressed exponentially as $J_{\text{eff}}\sim V e^{-l/\xi}$. Therefore the quantum state of the $N$ spins can be approximately revived with the spin echo protocol applied at the same time to all the designated q-bits, if the wait time is smaller than $1/J_{\text{eff}}$. Thus the buffer zone provided by the many-body localized medium provides inherent protection. Of course one should worry in this case from loss of fidelity due to the mismatch between the constituent and dressed q-bits which would be suppressed exponentially in the number of spins. It is possible that the passive protection provided by the medium should be supplemented by an active error correcting code to counter residual errors.

We wish to emphasize that storing quantum information in selected spins of the many-body localized medium is of course not claimed to be better than storing it in completely decoupled q-bits. Rather we have in mind a realistic situation, where the main source of decoherence of intended physical q-bits is their interaction with unwanted environment spins. For example, the main source of decoherence when using nitrogen-vacancy (NV) centers in diamond for purposes of quantum information is the interaction with randomly distributed $^{13}C$ nuclear spins. The latter also interact among themselves through dipolar interactions. The basic idea is that if these impurity spins were to form a many-body localized state (e.g. by effectively reducing the dimensionality \cite{Yao2013}) then their effect would be less disruptive.

\subsection{Topological states}

The MBL states discussed so far are in some sense extremely simple because they are perturbatively connected to states of particles fully localized to disjoint sites. Correspondingly the eigenstates in those systems can be mapped with arbitrary accuracy to simple product states through a finite depth unitary transformation\cite{Bauer2013}. But not all localized states must be so simple. In particular, topological order\cite{Huse2013} as well as symmetry protected topological states (SPT)\cite{Bahri2013,Chandran2014} can exist at high energy densities if the system is localized. In turn, such quantum order gives rise to coherent edge or defect states with dramatic dynamical manifestations. As we show below, measurements of the coherence properties of single q-bit like degrees of freedom can serve as sharp diagnostics, or order parameters, for the underlying topological properties.

The simplest way to see the establishment of topologically ordered or SPT states {\em at high energies} is to start with exactly solvable models, such as  Kitaev's Toric code\cite{Kitaev2003}. Such models, written only in terms of a complete set of  commuting operators remain exactly solvable also when the couplings are random. In this limit the entire spectrum has the same topological properties as the ground state, which stem from the structure of the integrals of motion. Weak interactions which break the exact solvability retain this structure.

%
The dynamical implications of topological structure in high energy eigenstates was demonstrated using a solvable spin-chain model, which realizes a many-body localized SPT phase with $Z_2\times Z_2$ symmetry \cite{Bahri2013}. In this case one can identify, on each edge of an open chain, a triplet of integrals of motion, which among themselves form a spin-1/2 algebra, i.e. $[\S^\a_{L},\S^\b_L]=i\e_{\a\b\g}\S^\g_L$ on the left edge and similarly on the right edge. The degrees of freedom represented by these operators therefore behave as {\em free} spin-$1/2$ degrees of freedom localized at the two edges. The transformations they generate between the degenerate edge states realize a projective representation of the symmetry group $Z_2\times Z_2$ \cite{Chen2011,Turner2013}.

Weak interactions, which spoil the exact solvability of the model
 cannot change the way symmetry is realized on the edge from a projective to a linear representation of the group, provided they respect the $Z_2\times Z_2$ symmetry of the model. Hence the free spin-$1/2$ edge states are protected  here through the same mechanism which protects the edge modes in gapped SPT ground states  \cite{Pollmann2010,Gu2009}. The important new element of course is that in many-body localized states the protection is not limited to the ground state degeneracy but extends to the entire spectrum.

It is quite remarkable that an emergent spin-1/2 degree of freedom at the edge is completely decoupled from all bulk degrees of freedom, while the constituent degrees of freedom are all fully interacting. Moreover this decoupling occurs without fine tuning, so long as the global protecting symmetry is present. A consequence of this fact, is that if we prepare the bare (strictly local) edge operator of a semi-infinite system in a state with an expectation value of any spin component $\av{\S^\a_L}$, it will  retain a non-vanishing expectation value in this orientation forever, regardless of how the state of the bulk was prepared. This is in sharp contrast to the integrals of motion discussed in the context of simple MBL states, where only one spin component, either $\tau^z$ or $\tau^x$ was conserved.

In a finite system of length $L$ interactions between the edge degrees of freedom will be generated due to the exponentially decaying  tails of the dressed edge operators $\tS^\a_{L,R}$. This would lead to  decay of $\av{S^\a_L(t)}$ on a time scale which grows exponentially with the system size, as was indeed demonstrated numerically\cite{Bahri2013}. Interestingly however, {\bf there is a qualitative difference between the decay of the edge spin in a system which can be mapped to free fermions and one with generic interactions}. In the former case, the only interactions generated between the two edges are ones which can be written as a quadratic fermion term and only two of such terms are allowed by the $Z_2\times Z_2$ symmetry. Therefore in the "non interacting" case the edge spin would, rather than decay, undergo oscillations beating between two frequencies. In the interacting case a huge number of coupling terms between the edges are generated, including strings with any number of bulk fermions connecting the two edges. In this case the oscillations of the edge spin will dephase, leading to actual decay of the expectation value $\av{S^\a_L(t)}$ on a time which gets exponentially bigger with system size.

In this section we considered the quantum coherence of the edge q-bits as a sharp diagnostic of the topological state. A more ambitious goal would be to use  MBL states as platforms for topological quantum computing. The standard vision of topological quantum computation \cite{Nayak2008} is to utilize non-abelian states (e.g. certain quantum Hall states). The quasiparticles of these states can in principle be used to store q-bits in a non local way, while unitary operations are performed by braiding quasi-particles. The topological protection is strictly in force only at zero temperature when there are no stray quasiparticles in the system. The actual error rate is therefore set by the temperature and is exponentially suppressed at low temperature due to presence of a gap.
Many body localized topological states can potentially provide much stronger protection. It has been pointed out that even if unwanted quasiparticles are excited, they can be harmless if they are localized by disorder\cite{Wootton2011,Stark2011}.

\section{Open questions}

Having surveyed the significant advances made during the last few years in understanding many-body localization we turn to outline what we see as some of the outstanding open questions. We note that there is some overlap with the corresponding survey of open questions given in another review appearing in this volume\cite{Nandkishore2014}

\subsection{The many-body localization transition}
In spite of the progress in describing the localized phase there is still very little understanding of the many-body localization transition. A transition from the many-body localized to a delocalized phase can be tuned, as originally conceived by Basko et. al\cite{Basko2006}, by tuning the energy across a many body (extensive) mobility edge.  Alternatively it may be  driven by changing the disorder or interaction strength at a constant energy density. Assuming this is a continuous phase transition, then it presents a completely new kind of critical point  in that it
must be accompanied by a fundamental change in the entanglement properties of the eigenstates: from area-law entanglement entropy in the localized eigenstates to volume law in the delocalized thermal eigenstates. This is very different both compared to classical (thermal) critical points and to quantum critical points.

As discussed in section \ref{sec:MBL-Comp} the area law entanglement in localized states is what made them amenable to theoretical and numerical analysis. The very processes which lead to delocalization also invalidate the RG approach developed to describe the localized phase and hinder efficient numerical simulation using matrix product states. Therefore, studies of the localization phase transition have so far been largely confined to brute force exact diagonalization of very small systems \cite{Oganesyan2007,Pal2010,Kjall2014}. Here we should emphasize again that the interesting physics takes place in high energy eigenstates and requires full diagonalization of the system, which can be done only for up to 18 spins. In spite of the small size Kjall et. al. \cite{Kjall2014} were recently able to obtain reasonable finite size scaling collapse of numerical data by using the variance of the entanglement entropy as a scaling variable. These results support the case for a continuous phase transition exhibiting some form of universality.

A recent theoretical paper by Grover\cite{Grover2014} used the strong sub-additivity property of the entanglement entropy to derive stringent constraints on how the scaling of entanglement entropy with sub-system size can change in a continuous phase transition. Specifically, grover argued that in a direct transition to a thermal state the critical point itself must be thermal in the sense that the entanglement entropy should follow a volume law with the same pre-factor (specific entropy) as in the corresponding thermal state.
This paper left open the intriguing alternative possibility of having a non ergodic yet delocalized intermediate phase between the localized and fully thermal phase, which would exhibit a sub thermal volume law entanglement\cite{DeLuca2014}. In either case It remains an open challenge to characterize the  critical point which controls energy transport, entanglement propagation and dynamic response functions in the vicinity of the localization transition.


\subsection{Many-body localization in translationally invariant systems}

Our discussion of the many-body localized phase relied on the presence of a quenched disorder potential. However, it is a very intriguing question if some form of localized or non thermalizing {\em phase} (i.e. not a fine tuned integrable point) can occur in translationally invariant systems. In this regard it is interesting to note that the quenched random potential in real amorphous solids is in fact provided by the  underlying nuclei. The real microscopic Hamiltonian including both the electrons and the nuclei is in fact translationally invariant.

With this in mind, recent papers have studied the possibility of many-body localization in lattice models with two species of interacting particles\cite{Schiulaz2013,DeRoeck2013}. If one species is infinitely heavy, that is, has vanishing hoping amplitude, then in every eigenstate the positions of the particles of this species are fixed. In this case the heavy particles only provide a disorder potential on which the light particles localize.
Perturbative arguments as well as some numerics suggest that such localized phases may be stable toward including weak hopping of the heavy particles on the lattice\cite{Schiulaz2013,DeRoeck2013}. This phase would have essentially the same properties as the disordered MBL phases we discussed here.

A rather different and intriguing possibility for a non thermalizing phase was suggested by Grover and Fisher \cite{Grover2013}, who considered systems of heavy nuclei and light electrons in the continuum at high energies. In this case one can imagine a situation in which the particles are delocalized with respect to the lab frame, however electrons are localized with-respect to the instantaneous ion coordinates. That is, if we were to take a snapshot measurement of the ion positions, then the remaining electron wave function following such a projection  would be localized. So, while the state is characterized by volume law entanglement entropy, the entanglement entropy of the many-electron wave function after measurement of the ion coordinates would follow an area law. The new state, termed by  Grover and Fisher a "quantum disentangled liquid", if exists, would be an example of a delocalized yet non-ergodic phase.

All the ideas on localization without quenched disorder remain rather speculative at the moment. Numerical studies to test them are limited to very small size exact diagonalizations, which therefore lead to indefinite conclusions. At the same time analytic approaches to MBL states, such as the RG method discussed in section \ref{RG-eigenstates} make explicit use of the quenched randomness. New theoretical or numerical approaches are clearly needed to make further progress in this area.

\subsection{Dynamics in localized states with mobility edges}
The RG approach described in section \ref{RG-eigenstates} and the fixed point Hamiltonian written in terms of a complete set of quasi-local integrals of motion allow a simple description of the dynamics in the localized phase. However, this picture is strictly valid only when the entire many-body spectrum is localized.

In systems with a mobility edge we clearly cannot have a complete set of local integrals of motion.
However we should still be able to construct all states below the mobility edge as mutual eigenstates
of a more restricted set of local conserved quantities, with non zero matrix elements only in the localized subspace. It remains an open question how to construct and characterize such a restricted set of integrals of motion and to understand the implications for dynamics in systems with a many-body mobility edge.


\subsection{Experimental tests of many-body localization}

Perhaps the most important challenge with regard to many-body localization is to identify physical systems where this phenomenon can be demonstrated and studied experimentally.
We hope to have conveyed the fact that many-body localization is not a fine tuned point in parameter space, but  rather a whole dynamical phase of matter, with quite dramatic observable consequences. Nonetheless this phenomenon has not yet been demonstrated and studied experimentally. The difficulty stems from the requirement that the system should be decoupled from a bath of delocalized excitations. A requirement that is generally not met by conventional solids. Nonetheless we argue that there are at least two realistic routes to observe many-body localization phenomena. The obvious route is to use systems such as ultra-cold atomic gases, which are naturally decoupled from the environment. The second approach is to look for signatures of many-body localization in the AC response (or at finite times) in solids with weak coupling to a phonon bath.

In systems of cold atoms it is quite natural to prepare simple initial states and observe their evolution in real time. As a first test of many-body localization, an experiment of this type should identify local observables, such as density fluctuations, which do not relax to their equilibrium value. Second one would like to show that this saturation is robust to adding interactions. For a more detailed study it would be important to demonstrate, through echo protocols, that a certain amount of local coherence persists in the system. This would distinguish the system from a classical glass. Although several experiments with weakly interacting disordered systems have been done to demonstrate Anderson localization \cite{Roati2008,Billy2008,Kondov2011}, to our knowledge no systematic tests of many-body localization have been carried out so far.

Conventional solids in contrast to cold atomic gases are inevitably coupled to a phonon bath and therefore cannot be strictly localized. Nonetheless we argue that clear signatures of many-body localization phenomena can be investigated also in these systems too. The idea is to focus on the low frequency response of the system at arbitrary temperatures. If the system has a small relaxation rate $\Gamma$ with the bath, then there may be a wide range of frequencies $\Gamma\ll\w\ll T$, over which universal signatures of localized phases and phase transitions can be seen. For example, as explained in section \ref{sec:qcp}, such probes would allow to observe dynamical critical points separating different localized phases. Such transitions would show up as universal crossovers with $\Gamma$ providing the only scale  \cite{Pekker2014}. In addition to phenomena in the many-body localized phase, clear signatures of the many-body localization transition have been predicted in systems with weak coupling to a phonon bath\cite{Basko2007,Gopalakrishnan2014}.

In light of the new insights concerning many-body localization, it would be interesting to revisit experiments done with certain types of disordered many-body systems. Good candidates, for example, are disordered films which show a quantum superconductor insulator transition at zero temperature. In certain films it had even been established that the phonons become effectively decoupled from the electrons at very low temperatures\cite{Altshuler2009,Ovadia2009} making many-body localization all the more relevant. We have already mentioned NV centers in diamond, coupled to the $^{13}$C nuclear spins as another candidate material. This system benefits from having high level of control, allowing  to detect an manipulate individual spins. Finally it would be very interesting to revisit AC susceptibility measurements done in highly disordered magnets, such as LiHo$_x$Y$_{1-x}$F$_4$, which exhibits a quantum critical point as well as a spin-glass phase\cite{Wu1991}.

\section{Conclusions}

We have reviewed the recent progress made in understanding the dynamics in many-body localized systems. The natural setting for the discussion were closed quantum systems with non-vanishing and even high energy density, where eigenstates of thermalizing systems would show volume law entanglement. In contrast eigenstates of many-body localized systems exhibit area-law entanglement akin to quantum ground states. We have shown how this feature of the localized phase enables efficient computation and facilitates the use of powerful theoretical approaches to describe the dynamics. In particular, we reviewed a renormalization group approach, useful for capturing universal dynamics controlled by critical points which separate distinct localized phases.

A simple phenomenological description in terms of local conserved quantities emerges naturally from the RG approach.
Although the integrals of motion are not strictly local, we have shown that they are addressable, and quantum coherence stored in a single integral of motion can be retrieved, after arbitrary long time, in a spin echo protocol applied to strictly local degrees of freedom. The persistence of quantum coherence in local degrees of freedom can be regarded as a direct test of many-body localization, which would distinguish this phase from a classical glass.

It is important for us to emphasize that in spite of the significant progress, made over the last few years, understanding of some of the basic issues is still very much lacking. For example, there is still title understanding of the critical point, or even if there is a critical point, which controls the many-body localization transition. It is also not clear if localization is the only way to establish non thermalizing phases in a generic way or if there are other intriguing possibilities which do not involve localization in real space. To gain deeper insight into these issues may require the development of novel theoretical approaches. Finally, and most importantly, physical systems which would allow systematic experimental investigation of many-body localization phenomena need to be identified.

\section{Acknowledgements}
We thank Dimtry Abanin, Anushya Chandran, Rahul Nandkishore, Mikhail Lukin, Zlatco Papic, Maxym Serbyn, David Huse, Eugene Demler, Anatoli Polkovnikov, Erez Berg, Gil Refael, David Pekker, Vadim Oganesyan, Sid Parameswaran, Ashvin Vishwanath, Yasaman Bahri, Immanuel Bloch, Mark Fischer, Matthew Fisher, Achim Rosch and Joel Moore for useful discussions on the topic and/or related collaborations.

\bibliographystyle{phd-url-notitle}

\end{document}